\RequirePackage{fix-cm}
\documentclass[smallextended]{svjour3}       
\smartqed  
\usepackage{graphicx}
\usepackage{mathptmx}
\usepackage{amsmath,bm}
\usepackage{amsfonts}
\usepackage{amssymb}
\usepackage{graphicx}
\usepackage{multirow}
\usepackage{subcaption}
\usepackage{units}
\usepackage{siunitx}
\usepackage{url}
\usepackage{float}

\journalname{Experimental Astronomy}
\begin{document}

\title{Extreme Universe Space Observatory on a Super Pressure Balloon~1 calibration:
	from the laboratory to the desert}

\author{\small
	J.H.~Adams Jr.$^{a}$,
	L.~Allen$^{b}$,
	R.~Bachman$^{c}$,
	S.~Bacholle$^{c}$,
	P.~Barrillon$^{d}$,
	J.~Bayer$^{e}$,
	M.~Bertaina$^{f,g}$,
	C.~Blaksley$^{h}$,
	S.~Blin-Bondil$^{i}$,
	F.~Cafagna$^{j}$,
	D.~Campana$^{k}$,
	M.~Casolino$^{l,m}$,
	M.J.~Christl$^{n}$,
	A.~Cummings$^{c}$,
	S.~Dagoret-Campagne$^{d}$,
	A.~Diaz Damian$^{o}$,
	A.~Ebersoldt$^{p}$
	T.~Ebisuzaki$^{l}$,
	J.~Escobar$^{q}$,
	J.~Eser$^{c,*}$,
	J.~Evrard$^{r}$,
	F.~Fenu$^{f,g}$,
	W.~Finch$^{c}$,
	C.~Fornaro$^{s}$,
	P.~Gorodetzky$^{h}$,
	R.~Gregg$^{c}$,
	F.~Guarino$^{k,t}$,
	A.~Haungs$^{p}$,
	W.~Hedber$^{q}$,
	P.~Hunt$^{c}$,
	A.~Jung$^{h}$,
	Y.~Kawasaki$^{l}$,
	M.~Kleifges$^{p}$,
	E.~Kuznetsov$^{a}$,
	S.~Mackovjak$^{u}$,
	L.~Marcelli$^{m}$,
	W.~Marsza{\l }$^{v}$,
	G.~Medina-Tanco$^{q}$,
	S.S.~Meyer$^{b}$,
	H.~Miyamoto$^{f,g}$,
	M.~Mastafa$^{a}$,
	A.V.~Olinto$^{b}$,
	G.~Osteria$^{k}$,
	W.~Painter$^{p}$,
	B.~Panico$^{k}$,
	E.~Parizot$^{h}$,
	T.~Paul$^{w}$,
	F.~Perfetto$^{k}$,
	P.~Picozza$^{m,x,l}$,
	L.W.~Piotrowski$^{l}$,
	Z.~Plebaniak$^{v}$,
	Z.~Polonski$^{c}$,
	G.~Pr\'ev\^ot$^{h}$,
	M.~Przybylak$^{v}$,
	M.~Rezazadeh$^{b}$,
	M.~Ricci$^{y}$,
	J.C.~Sanchez~Balanzar$^{q}$,
	A.~Santangelo$^{e}$,
	F.~Sarazin$^{c}$,
	V.~Scotti$^{k,t}$,
	K.~Shinozaki$^{f,g,v}$,
	J.~Szabelski$^{v}$,
	Y.~Takizawa$^{l}$,
	L.~Wiencke$^{c}$,
	R.~Young$^{n}$,
	P.~von Ballmoos$^{o}$
}

\authorrunning{J.H.~Adams~Jr.~et~al.} 

\institute{
* corresponding author\\
$^{a}$ University of Alabama in Huntsville, Huntsville, USA\\
$^{b}$ University of Chicago, USA\\
$^{c}$ Colorado School of Mines, Golden, USA\\
$^{d}$ LAL, Univ Paris-Sud, CNRS/IN2P3, Orsay, France\\
$^{e}$ Institute for Astronomy and Astrophysics, Kepler Center, University of T\"ubingen, Germany\\
$^{f}$ Istituto Nazionale di Fisica Nucleare - Sezione di Torino, Italy\\
$^{g}$ Dipartimento di Fisica, Universita' di Torino, Italy\\
$^{h}$ APC, Univ Paris Diderot, CNRS/IN2P3, CEA/Irfu, Obs de Paris, Sorbonne Paris Cit\'e, France\\
$^{i}$ Omega, Ecole Polytechnique, CNRS/IN2P3, Palaiseau, France\\
$^{j}$ Istituto Nazionale di Fisica Nucleare - Sezione di Bari, Italy\\
$^{k}$ Istituto Nazionale di Fisica Nucleare - Sezione di Napoli, Italy\\
$^{l}$ RIKEN, Wako, Japan\\
$^{m}$ Istituto Nazionale di Fisica Nucleare - Sezione di Roma Tor Vergata, Italy\\
$^{n}$ NASA - Marshall Space Flight Center, USA\\
$^{o}$ IRAP, Universit\'e de Toulouse, CNRS, Toulouse, France\\
$^{p}$ Karlsruhe Institute of Technology (KIT), Germany\\
$^{q}$ Universidad Nacional Aut\'onoma de M\'exico (UNAM), Mexico\\
$^{r}$ CNES, Toulouse, France\\
$^{s}$ Uninettuno University, Rome, Italy\\
$^{t}$ Universita' di Napoli Federico II - Dipartimento di Scienze Fisiche, Italy\\
$^{u}$ Institute of Experimental Physics, Kosice, Slovakia\\
$^{v}$ National Centre for Nuclear Research, Lodz, Poland\\
$^{w}$ Lehman College, City University of New York (CUNY), USA\\
$^{x}$ Universita' di Roma Tor Vergata - Dipartimento di Fisica, Roma, Italy\\
$^{y}$ Istituto Nazionale di Fisica Nucleare - Laboratori Nazionali di Frascati, Italy\\
}

\date{Received: date / Accepted: date}

\maketitle

\begin{abstract}
The {\it Extreme Universe Space Observatory on a Super Pressure Balloon~1} (EUSO-SPB1) instrument was launched out of Wanaka, New Zealand, by NASA in April, 2017 as a mission of opportunity. The detector was developed as part of the {\it Joint Experimental Missions for the Extreme Universe Space Observatory} (JEM-EUSO) program toward a space-based ultra-high energy cosmic ray (UHECR) telescope with the main objective to make the first observation of UHECRs via the fluorescence technique from suborbital space.
The EUSO-SPB1 instrument is a refractive telescope consisting of two \unit[1]{m$^2$} Fresnel lenses with a high-speed UV camera at the focal plane. The camera has 2304 individual pixels capable of single photoelectron counting with a time resolution of \unit[2.5]{$\mu$s}.
A detailed performance study including calibration was done on ground. We separately evaluated the properties of the Photo Detector Module (PDM) and the optical system in the laboratory. An end-to-end test of the instrument was performed during a field campaign in the West Desert in Utah, USA at the Telescope Array (TA) site in September 2016. The campaign lasted for 8 nights.
In this article we present the results of the preflight laboratory and field tests. Based on the tests performed in the field, it was determined that EUSO-SPB1 has a field of view of \unit[11.1]{$^\circ$} and an absolute photo-detection efficiency of 10\%. We also measured the light flux necessary to obtain a 50\% trigger efficiency using laser beams. These measurements were crucial for us to perform an accurate post flight event rate calculation to validate our cosmic ray search. Laser beams were also used to estimated the reconstruction angular resolution. Finally, we performed a flat field measurement in flight configuration at the launch site prior to the launch providing a uniformity of the focal surface better than 6\%.

\end{abstract}

\section{Introduction}
\label{sec:Intro}
\begin{figure}[h!] 
	\centering 
	\includegraphics[width = 1.\textwidth]{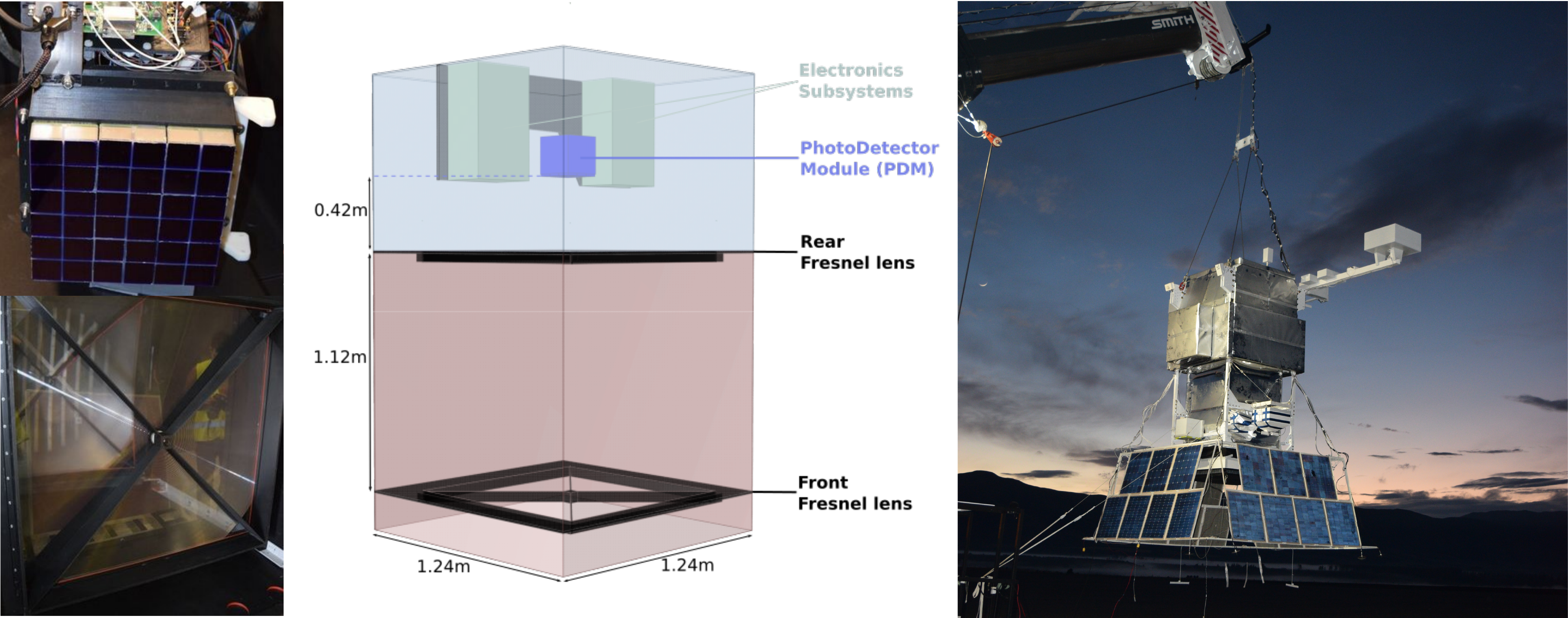}
	\caption{The EUSO-SPB1 instrument: The camera with the 2304 single photoelectron
		counting pixels and the lenses in the left panel, a sketch of arrangement of the key components (electronics, camera, and Fresnel lenses) shown in the center panel, and a picture in the right panel taken shortly before launch of the instrument as integration into balloon gondola system.}
	\label{fig:100_PDMandInstrument}
\end{figure}

The {\it Extreme Universe Space Observatory on a Super Pressure Balloon~1} (EUSO-SPB1, see Fig. \ref{fig:100_PDMandInstrument}) is a pathfinder experiment towards space-based optical cosmic ray telescopes. Such a space telescope will have a much larger observational volume than ground based telescopes to map the entire sky at ultra-high energies (E $>$ \unit[10$^{18}$]{eV}) and discover the still unknown sources of ultra-high energy cosmic rays (UHECR). The initial JEM-EUSO balloon mission, EUSO-Balloon, was an overnight flight in 2014 launched from Timmins, Canada. The detector measured UV backgrounds looking down. The camera readout was triggered by an internal clock. Through accidental coincidence the instrument also captured several hundred artificial tracks produced by a pulsed UV laser aboard a helicopter flown below the balloon. For more details on the EUSO-Balloon experiment, please check \cite{balloon_2015_Adams_Overview}. Following a successful recovery, the instrument was upgraded extensively for a long duration flight. The science goals for this mission, dubbed EUSO-SPB1, were to make the first observations of tracks from ultra high energy cosmic rays with a UV fluorescence detector looking down on the Earth, measure UV backgrounds over ocean and over clouds, and search for fast pulses of light from other processes. Raising the technology readiness level of camera elements for space missions was a primary technology objective.
\par
Briefly the EUSO-SPB1 detector consisted of two \SI{1}{\square\metre} PolyMethyl MethAcrylate (PMMA) Fresnel lenses \cite{JEMEUSO_2009_Takahashi_lens} serve as the optical system and focus the light from a source onto a Photo Detector Module (PDM) \cite{balloon_2015_Jung_PDM}. The PDM has a modular design: nine Elementary Cells (EC) each consisting of four Multi-Anode PhotoMultiplier Tubes (MAPMT, Hamamatsu R11265-113-M64-MOD2) with 64 individual channels. The total number of pixels is 2304 (48$\times$48) pixels. The pixel size is 2.88$\times$\unit[2.88]{mm$^2$}, equal to the active area. BG3 UV bandpass filters (\unit[2]{mm} thickness), glued on top of each MAPMT, mostly select light between 290 and \SI{430}{\nm} to observe air fluorescence photons. The instrument operates in single photoelectron counting mode with a time resolution of \SI{2.5}{\micro\second}; we define this time as 1 Gate Time Unit (GTU).
A detailed description of the detector can be found in Ref. \cite{SPB1_2017_Bacholle_Instrument}. The data processing system handles the data acquisition and the communication with the ground via the telemetry system provided by NASA \cite{SPB1_2019_Scotti_DP,SPB1_2019_Fornaro_DP}.\\
NASA launched EUSO-SPB1 from Wanaka, New Zealand on April 25, 2017 as a mission of opportunity. The flight lasted for 12 days and \SI{4.5}{\hour} before it had to be terminated above the Pacific Ocean. Shortly into the flight the balloon lost its super pressure state due to damage of the balloon envelope, making a long duration flight impossible. In total the detector recorded data for around \SI{40}{\hour}. A full description of the mission is given in Ref. \cite{SPB1_2017_Wiencke_overview} \cite{SPB1_Journal}.\\
Before the launch a full characterization of the instrument was performed. The calibration had two major parts, the laboratory measurements for single components and the field tests for an end-to-end characterization.
The field tests were necessary to achieve a full-scale test of the instrument and to estimate its capability of measuring tracks from an Extensive Air Shower (EAS) induced by UHECR. This was also the only way to estimate the energy threshold of the instrument. A detailed calibration prior to the flight was crucial due to the typical risk of not being able to recover the payload.
\section{Laboratory Tests}
\label{sec:LabTest}
The preflight test and calibration of the instrument was divided in two phases. In the first phase we performed tests in a laboratory environment of most of the single components of the detector, mainly the PDM and the optical system (discussed in this section). After these tests we assembled the entire instrument, transported it to the field (see section \ref{sec:transport}) and performed a full end-to-end characterization there as described in Section \ref{sec:FieldTest}.

\subsection{Calibration of the PDM}
\label{subsec:CalPDM}
\begin{figure}[h]
	\centering
	\begin{subfigure}{.48\textwidth}
	  \centering
	  \includegraphics[width=\textwidth]{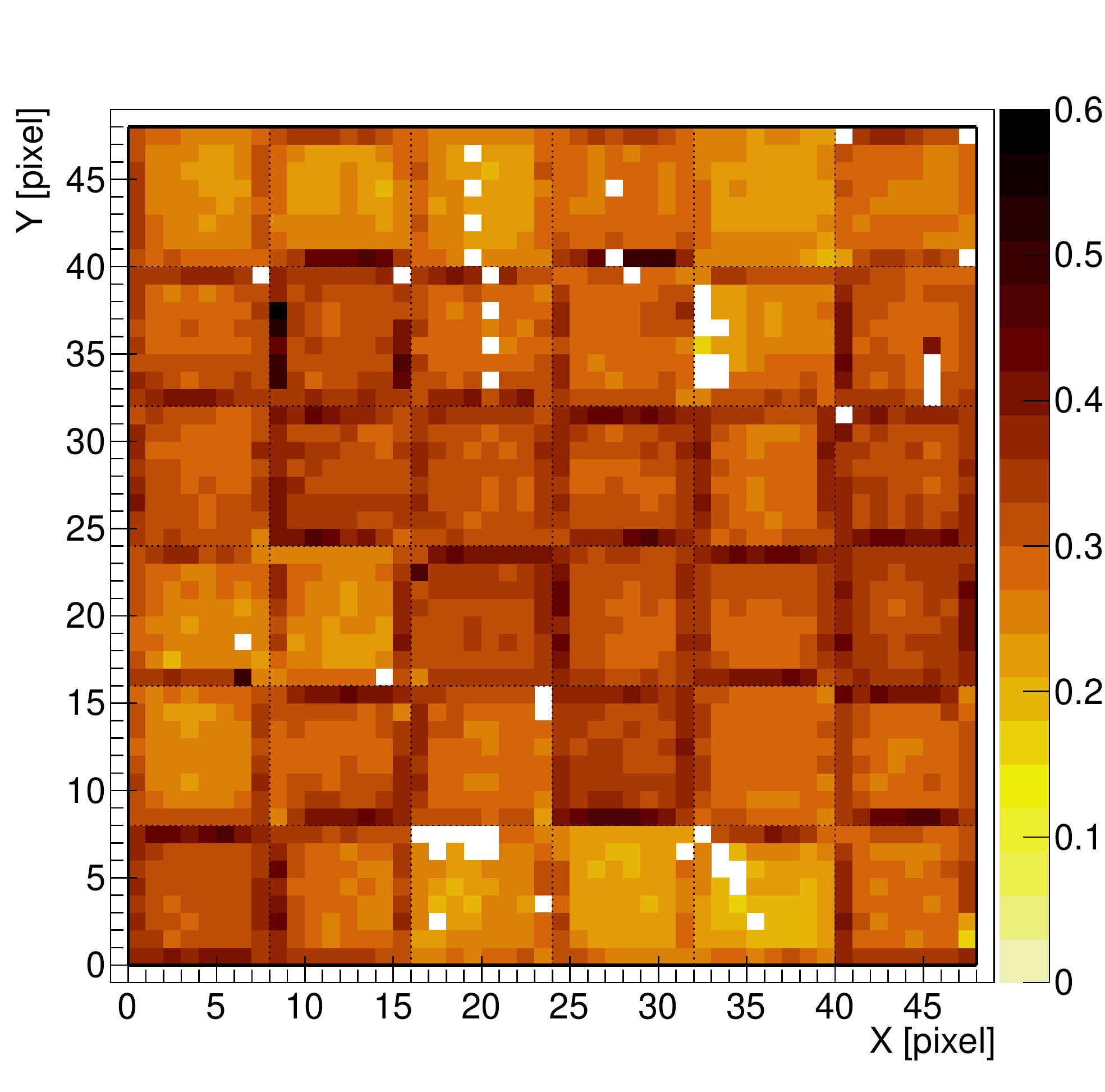}
	  \caption{}
	  \label{fig:AbsExample}
	\end{subfigure}%
	\hfill
	\begin{subfigure}{.48\textwidth}
	  \centering
	  \includegraphics[width=\textwidth]{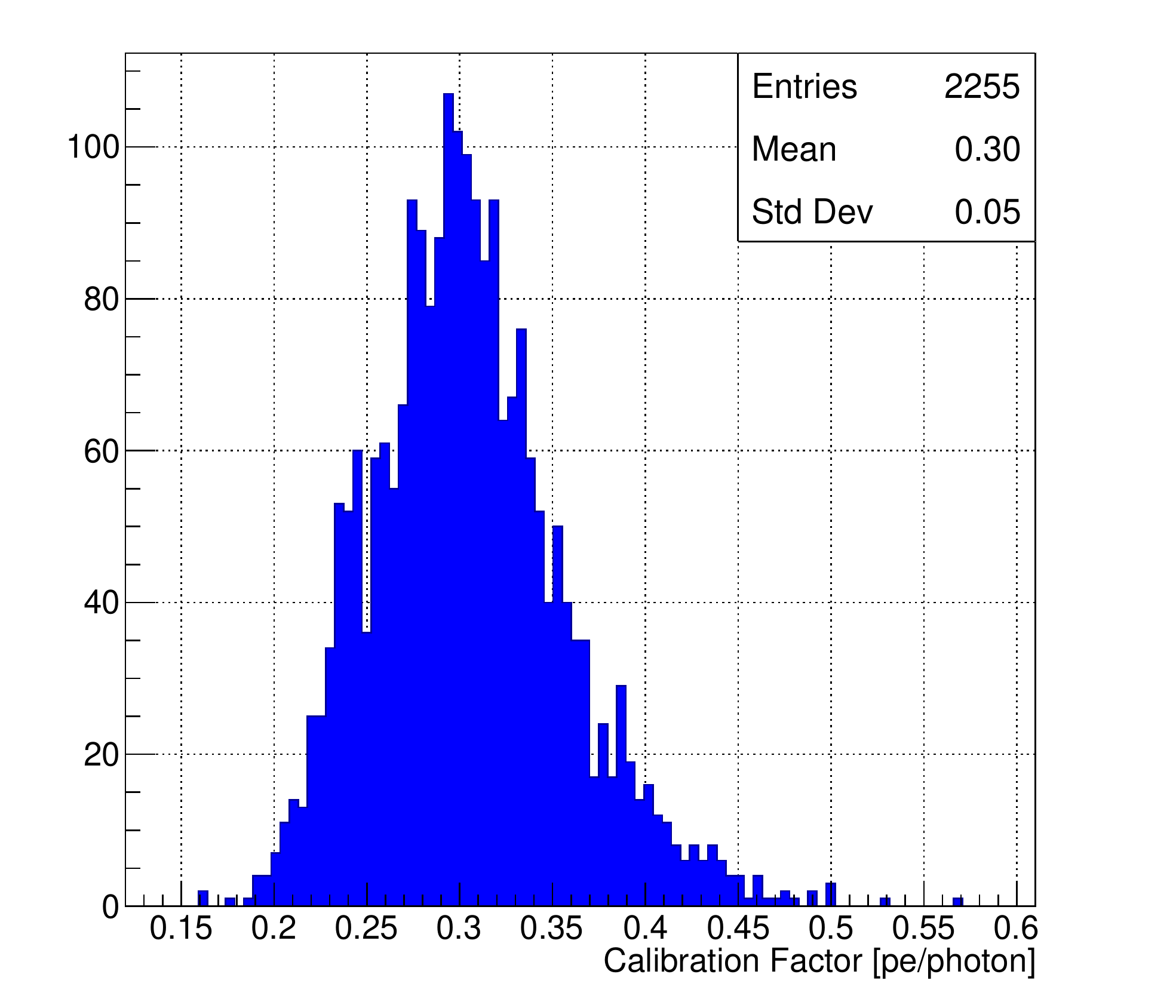}
	  \caption{}
	  \label{fig:labAbsResult}
	\end{subfigure}
	\caption{Absolute calibration of the PDM performed at the APC laboratory, with a \SI{378}{\nm}-LED. (a) Example frame of the uniformly illuminated PDM at the standard setting; color code is calibration factor with white indicating non functioning pixels. (b) Histogram of the calibration factor (photoelectron(pe) per photon) for each pixel of the focal surface.}
	\label{fig:labCal}
\end{figure}

An absolute photometric calibration provides a relationship between the amount of UV-photons arriving at the detector and the measured signal. At the Astroparticle and Cosmology (APC) laboratory in Paris (France), we carried out relative and absolute photometric calibrations of the PDM of EUSO-SPB1. An in-depth description of the setup is given in Ref. \cite{balloon_2014_Blaksley_PMTCal}.
We aligned the port of a 4 inch integrating sphere closed by a 3 mm diameter diaphragm from a distance of \SI{49}{\cm} to the center of one of the 36 MAPMTs. The light output is considered Lambertian. The sphere was equipped with 2 more ports, one for the light injection (\SI{378}{\nm} LED), and one for a NIST calibrated photodiode \cite{OphirPD}, used to monitor the light output of the sphere. The difference in light intensity over the surface of the MAPMT was less than 1 \%. The process was repeated for all 36 PMTs, and from the obtained data we computed the relative efficiency of each pixel of the full PDM.\\
For one of the above measurements, a calibrated photodiode \cite{OphirPD} measured the light flux at the illuminated PMT position for the purpose of absolute calibration. 
In the next step we compared the count on each pixel, normalized based on the relative calibration, to the absolute intensity measured by the photodiode at the illuminated PMT position. The ratio of these values are the absolute calibration factors that describe how many counts are produced by one photon incident on the BG3 filters. The calibration factor is the combined effect of three efficiencies at the relevant wavelength: the quantum efficiency of the PMT (37\%),  the photoelectron collection efficiency (around 80\%) and the transmission of the BG3 filter ($>$90\%). This is displayed in Fig. \ref{fig:AbsExample} where the color codes the calibration factor for the PDM. Fig. \ref{fig:labAbsResult} shows the histogram of the calibration factors for the pixels of the camera. The number of entries reflects the number of operating pixels in the camera. The apparent higher efficiency observed at two of the borders of each MAPMT is related to a manufacturing issue that was first identified with these measurements and contributes to the spread of the calibration factor. 
The dominant uncertainties of these measurements were the uncertainty to place the photodiode at the same distance from the LED as the BG3 filters of the PDM were as well as the uncertainty on the sensitive area of the photodiode. For all pixels, 1 photon produces \SI[separate-uncertainty = true,multi-part-units=single]{0.30(3)}{photoelectrons} (pe) on average. The quoted error is obtained by the uncertainties of the photodiode.

\subsection{Optics Characterization}
\label{subsec:CalLens}
\begin{figure}[h!] 
	\centering 
	\includegraphics[width = 1.\textwidth]{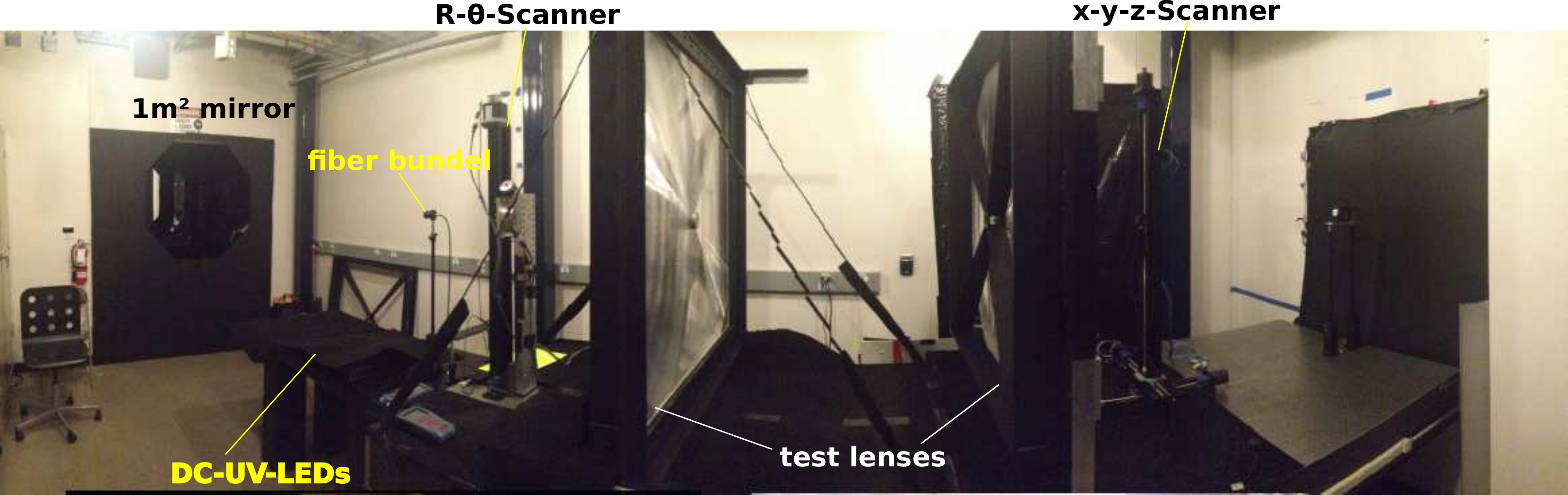}
	\caption{Test stand for optical system performance measurement as a setup for EUSO-SPB1 at the Colorado School of Mines in August 2016.}
	\label{fig:220_LensTestStand}
\end{figure}

\begin{figure}
	\centering
	\begin{subfigure}[b]{.45\textwidth}
	  \centering
	  \includegraphics[width=\linewidth]{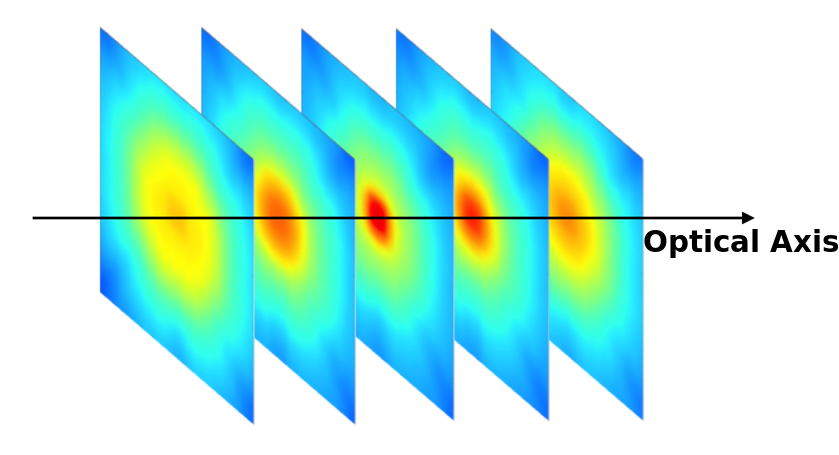}
	  \caption{}
	  \label{fig:220_CubeScan}
	\end{subfigure}%
	\hfill
	\begin{subfigure}[b]{.45\textwidth}
		\centering
		 \includegraphics[width=\linewidth]{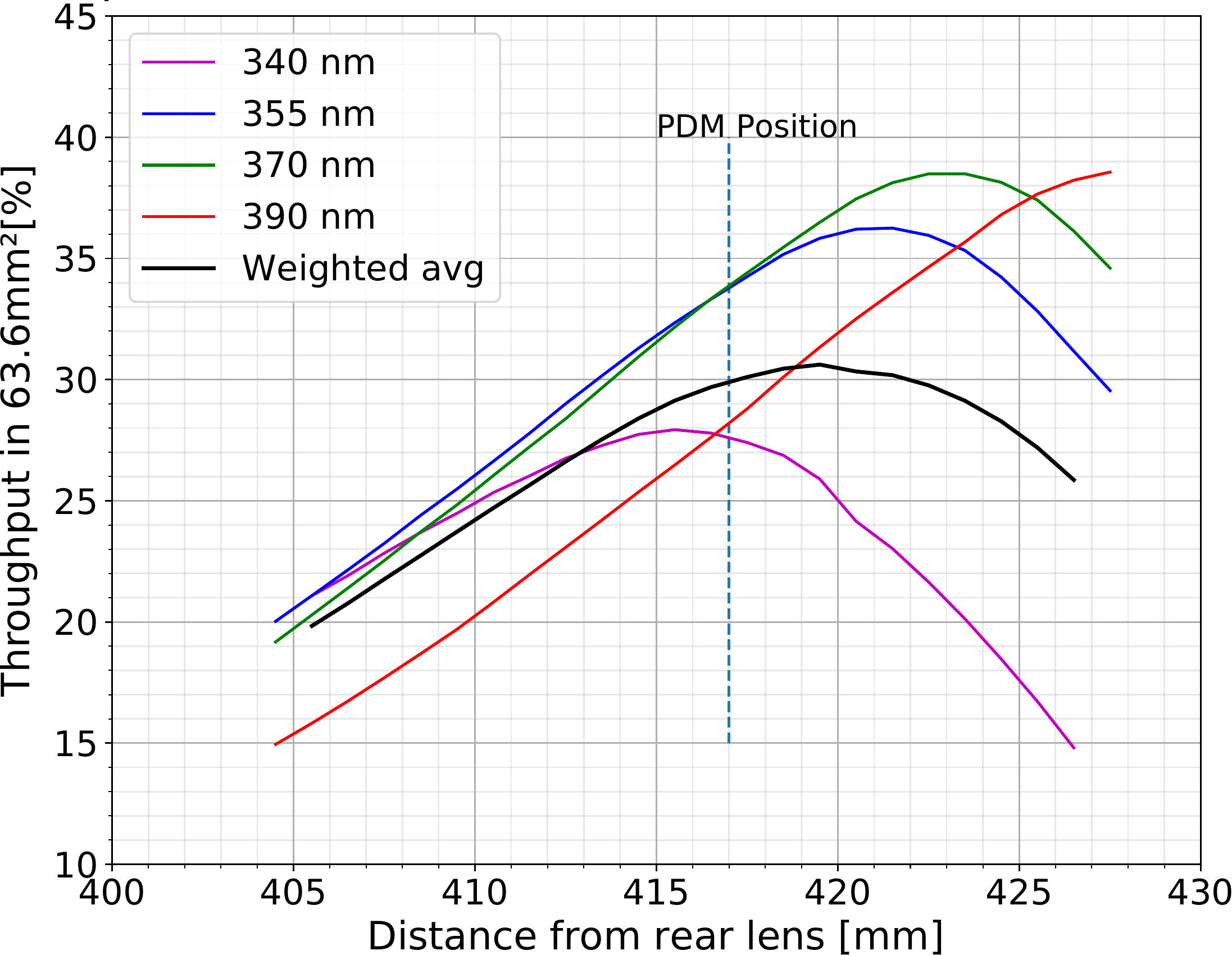}
		 \caption{}
		 \label{fig:220_CubeResult}
	\end{subfigure}
	\caption{Cube Scan: (a) shows a visualization of the scan. Each frame shows the light intensity (color coded) at different distances along the optical axis. (b) shows the result of the measured throughput for the 2-lens system a long the optical axis. The PSF was independently determined to be enclosed in an area of  \SI{63.6}{\square\milli\meter} (spot diameter is \SI{9}{\milli\meter}). The black line corresponds to a weighted average of the measurements based on the air fluorescence spectrum.}
	\label{fig:220_Cube}
\end{figure}

EUSO-SPB1 has a baseline configuration of two refractive Fresnel lenses. Another configuration (3 lenses) includes a diffractive middle lens, a prototype, for chromatic correction and optimization of the Point Spread Function (PSF). One important step in the characterization was to determine which of the two configurations would lead to a higher probability to detect EAS. We tested both systems first in a \SI{1}{\meter} diameter, parallel beam of UV light in a test stand under laboratory conditions (Fig. \ref{fig:220_LensTestStand}).
As a calibration source we coupled DC-UV LEDs (four different wavelengths) to a Y-fiber bundle. We placed one end of the fiber in the focal point of a diffraction limited \SI{1}{\meter} mirror; the other end was used to monitor the stability of the LEDs with a NIST traceable photodiode \cite{OphirPD}. The accuracy of the photodiode is the main source of the error related to these measurements and quoted by the manufacturer to be $\pm$ 3 \%. We needed the fiber setup to minimize obstruction of the reflected light. In this way, we were able to exchange the LEDs without risking changes in the alignment of the collimator. To obtain a narrow wavelength window around the target wavelengths of \SIlist{340;355;370;390}{\nano\meter}, we installed \SI{10}{\nano\meter} bandpass filters. The resulting parallel beam was sampled at 408 positions using a custom built R-$\theta$-``windmill" scanner, which allowed to determine the absolute flux incident on our Fresnel optics. The scanning interval for R was between \SIlist[list-units = single]{1;23}{\centi\meter} in \SI{2}{\centi\meter} steps. For $\theta$, we scanned from \SIrange{10}{350}{\degree} in \SI{10}{\degree} steps. The 2-and 3-lens configuration underwent three different tests. Each was performed at all four wavelengths: a scan along the optical axis (efficiency), a cube scan around the coarse predefined focal point (efficiency, PSF and wavelength dependency) and a slit scan perpendicular to the optical axis (PSF). During the cube scan test we took measurements forming a 3D grid which contains images of the PSF at different positions along the optical axis (Fig. \ref{fig:220_CubeScan}). The grid size for the fine scan around the predetermined focal point was \unit[3]{cm} in all dimensions and the step size was \unit[1]{mm}. To determine the PSF, we defined the area in which the light flux is at least 10\% of the value of the maximum signal within this grid. For the transmission efficiency, we define the ratio of the photon flux focused within \SI{63.6}{\square\milli\meter} (area of PSF) to that of the incident photon on the front lens (\unit[0.93]{m$^2$}). We show the result of the scans for all wavelengths in Fig. \ref{fig:220_CubeResult}. We also calculated an average weighted by the air fluorescence spectrum (black line) as given in Ref. \cite{2007_Ave_FluoEmission}.
The low flux of light transmitted through the lenses is caused by light scattering on microscopic oscillations in the surface profile of the lenses. These oscillations arose from the diamond turning process used to manufacture the lenses. The relatively low transmittance is the results of the combined effects on both lenses, as already seen in the EUSO-Balloon experiment \cite{balloon_2019_Diaz_lens_performens}. This explains in part the reason why the measured PSF and transmission efficiency do not match well the initial ray-tracing simulations performed prior to the production of the lenses.
Table \ref{tab:220_LensProperties} contains the final results of all the tests for both lens configurations. Although the 3-lens system reduces the aberration and improves the PSF - as expected - the loss in efficiency is too large (2 times lower than the 2-lens system). At the end, the amount of light collected on the focal spot remains the main driver for achieving the lowest possible energy threshold for EUSO-SPB1.
\begin{table}
	\centering
	\caption{Optics characterization results for the 2-lens system (flight configuration) and the 3-lens system (containing diffractive lens)}
	\label{tab:220_LensProperties}
	\begin{tabular}{p{3.2cm}|c|c}
		  & 2 lens system & 3 lens system\\ 
		 \hline Transmission efficiency & 31 $\pm$ 2\% & 15 $\pm$ 1\%\\
		 (within \SI{63.6}{\square\milli\meter}) & & \\ 
		 Spot size & \SI{9}{\milli\meter} diameter & \SI{2}{\milli\meter} diameter\\
		                   & $\sim$ 3$\times$3 pixel grid & $\sim$ 1$\times$1 pixel grid\\
		 Displacement in focal points of \SIlist[list-units = single]{340;390}{\nm} & \unit[15]{mm} & \unit[1]{mm}\\  
	\end{tabular} 
\end{table}
  
\subsection{Flat Fielding}
\label{subsec:FlatField}
\begin{figure}[htb] 
   \centering
   \begin{subfigure}[b]{.48\textwidth}
     \includegraphics[width=\textwidth]{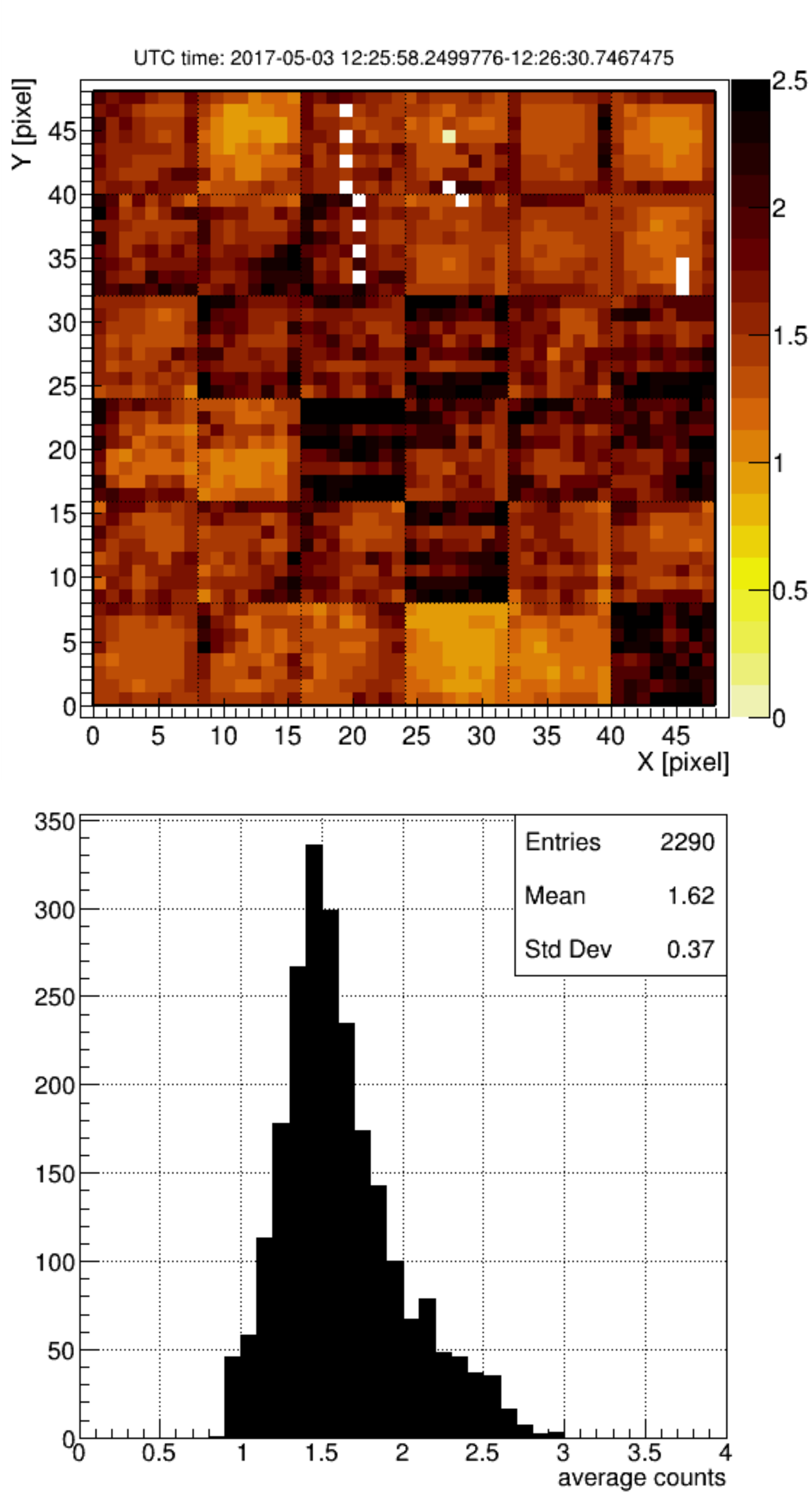}
     \caption{}
   \end{subfigure}
   \hfill
   \begin{subfigure}[b]{.48\textwidth}
     \includegraphics[width=\textwidth]{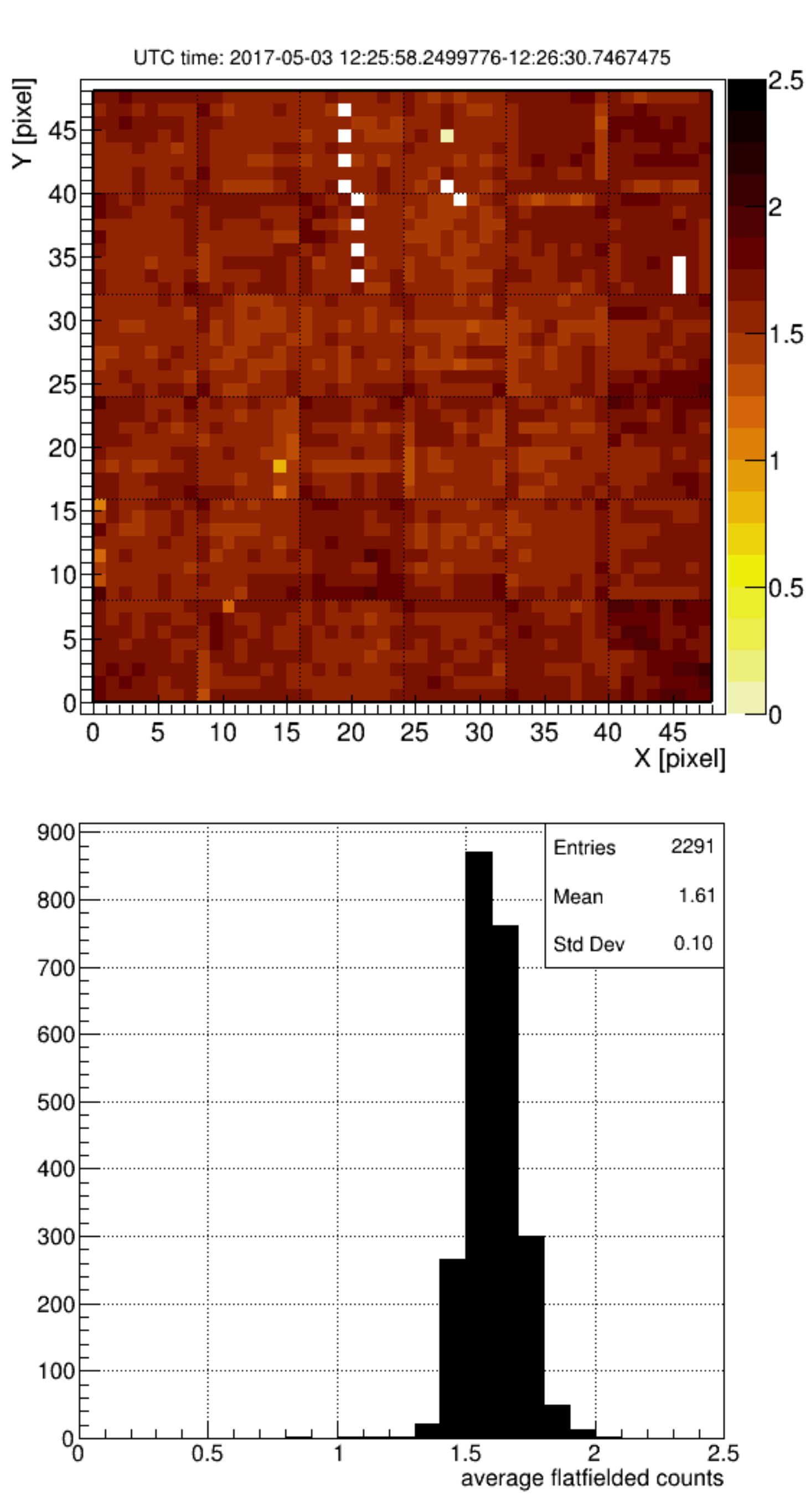}
     \caption{}
   \end{subfigure}
   \caption{The frame is an average of 2048 single frames (16 triggers 128 frames each) during the flight recorded on May 3, 2017 at 12:25:52 UTC. The color indicated the averaged photoelectron counts.
   The top two panels show a Background measurement during the flight. The bottom two panels show the corresponding histograms of the individual pixel response before and after flat fielding in (a) and (b), respectively.
   }
   \label{fig:230_FlatFielding}
\end{figure}

To compensate for non-uniform sensitivity across the PDM and for potential distortions in the optical path we used a flat field correction. We created a uniform signal on the PDM by illuminating a \unit[2.4]{m} by \unit[3.7]{m} wide screen covered with Tyvek\texttrademark1056 using a \SI{365}{\nano\meter} pulsed LED. We placed the screen \SI{4.65}{\meter} in front of the instrument aperture in a darkened room. The above geometry allowed us to illuminate the entire Field of View (FoV). The reflected light from the screen was diffuse: we thus assumed the signal on the focal surface was uniform. This is only true as our FoV is small enough that effects like vignetting are negligible. We estimated the factor that has to be applied to create a uniform response by normalizing the counts in each pixel to the average count over the full camera. To cover the linear part of the dynamic range (\unit[1 to about 30]{pe per pixel per GTU}) of the detector (assuring a uniform response over the whole range), we pulsed the LED at different intensities.\\
We repeated the same procedure prior to the launch in Wanaka (New Zealand) in flight configuration. By using a crane, we hung the instrument \SI{5}{\meter} above a by \unit[3.66]{m$^2$} Tyvek\texttrademark1025D screen on the ground. This allowed us to spin the detector around the optical axis accounting for potential irregularities in the diffuseness of the reflected light. Such irregularities could have come from surrounding light as this measurement was conducted outside and not in a laboratory. We show the effect of the flat fielding in Fig. \ref{fig:230_FlatFielding} on data obtained during the actual flight before (left) and after (right) flat fielding correction.
As can be seen, after flat fielding, the spread of the pixel sensitivity is around 6\%. The number of entries reflects the operating pixels of the camera. The flat fielding result obtained in Wanaka is the one we used in the later analysis of the flight data.
The remaining 6\% spread in pixel sensitivity is expected to some extent and not an indication of an inaccurate flat frame. It is mainly based due to the illumination source which is real emission of the ground and the atmosphere. These emissions are non-uniform by nature. In addition each of the 16 events fulfilled at least for a few frames the trigger condition (signal over threshold) meaning some pixel recorded a real signal.

\section{Transport of Assembled Instrument to Field Testing Location}
\label{sec:transport}

\begin{figure}[h!] 
	\centering 
	\includegraphics[width = 1.\textwidth]{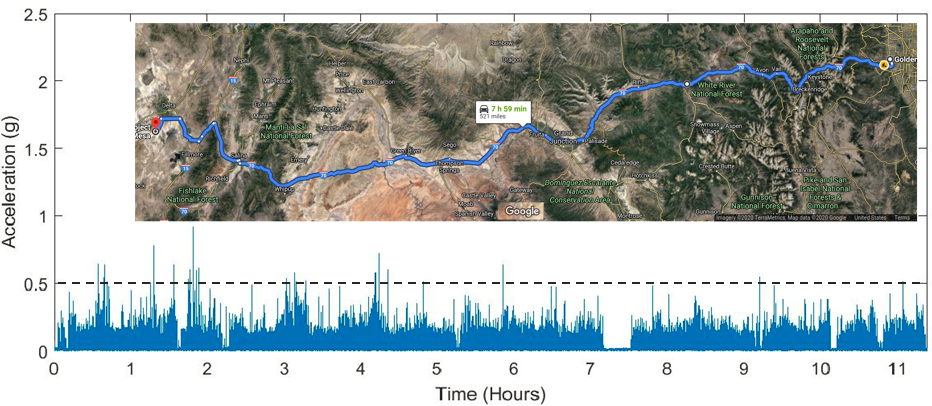}
	\caption{Data logged by an accelerometer that was attached to the assembled fluorescence telescope for the drive from the Golden, CO, assembly lab to the Utah desert for field tests \cite{SPB1_2017_Cummings_FieldTest} (\textit{created with Google maps, Imagery Terra Metrics, Map data Google)}. The dashed line at \unit[0.5]{$g$} represents a NASA requirement (see text). Higher spikes were typically encountered on highway bridge joints.}
	\label{fig:300_accelerometer_to_utah}
\end{figure}

Prior to qualification testing at the NASA facility in Palestine, Texas, we transported the fully assembled instrument and a roving UV laser system~\cite{2015_Hunt_GLS,Masterthesis_2015_Hunt_GLS} in two customized trailers from the assembly laboratory at the Colorado School of Mines in Golden, Colorado, to the Telescope Array \cite{TA} site in the West Desert in Utah for field tests. Preparations for this 1600~km round trip were extensive~\cite{SPB1_2017_Cummings_FieldTest} and included driving a dummy load over the planned route inside the air-ride trailer on top of the vibration damped custom dolly designed for moving the flight instrument to identify potential vibration hazards. For the actual transport we used a pilot vehicle to report potential dangerous road conditions (e.g. pot holes) to the team transporting the instrument. Data logged by the accelerometer mounted on the flight instrument during the trip to the desert was mostly below \unit[0.5]{$g$} with occasional bigger bumps (Fig.~\ref{fig:300_accelerometer_to_utah}). (For reference, NASA requires that flight ready balloon payloads must be able to tolerate jolts of \unit[0.5]{$g$} during the transfer to the launch pad.)
In the desert, the trailer also served as the field enclosure for EUSO-SPB1 and was 
positioned next to the EUSO-TA prototype \cite{eusota_2018_Piotrowski_results} for simultaneous observations. 
We tilted the dolly up to point the detector optical axis about 8${^\circ}$ above the horizon. Four hours after arriving in the desert, the EUSO-SPB1 detector turned on smoothly and started to record first tracks from lasers.
\section{Field Tests}
\label{sec:FieldTest}
The field campaign lasted for 8 nights during which we recorded hundreds of thousands of laser shots in different directions and at different energies provided by the roving laser system set up in a distance of \unit[24]{km} from the instrument in the center of the FoV. The scattered light of the laser beam produces a similar signature in our instrument as an EAS track, but we can control the energy and direction. This makes laser the ideal ``test beams" for optical cosmic ray detectors. 
This laser trailer is a prototype for a Global Light System (GLS) station \cite{2015_Hunt_GLS,2013_Wiencke_GLS}. As a light source we used a \SI{355}{\nm} frequency tripled Nd:YAG laser with the possibility to point in any direction above the horizon with a precision of \SI{0.2}{\degree}. Its energy ranges from \SI{250}{\micro\J} to \SI{80}{\m\J}. The beam is randomly polarized within 4$\%$ and the energy is monitored with an accuracy of 5$\%$. The system was tested during multiple earlier campaigns. Besides measuring the FoV with two independent methods, we also performed an absolute calibration and estimated the instrument energy threshold (50\% trigger efficiency) for the two different lens configurations (2 and 3-lens systems). Finally, we measured the angular resolution of EUSO-SPB1. The setup and results of all tests are presented in detail in the following section. Besides the laser and other calibration events, we also recorded background data including airplanes, stars and meteors.

\subsection{Field of View Measurements}
\label{subsec:FoV}
\begin{figure}[h]
  \centering
  \begin{subfigure}[b]{.48\textwidth}
    \includegraphics[width=\linewidth]{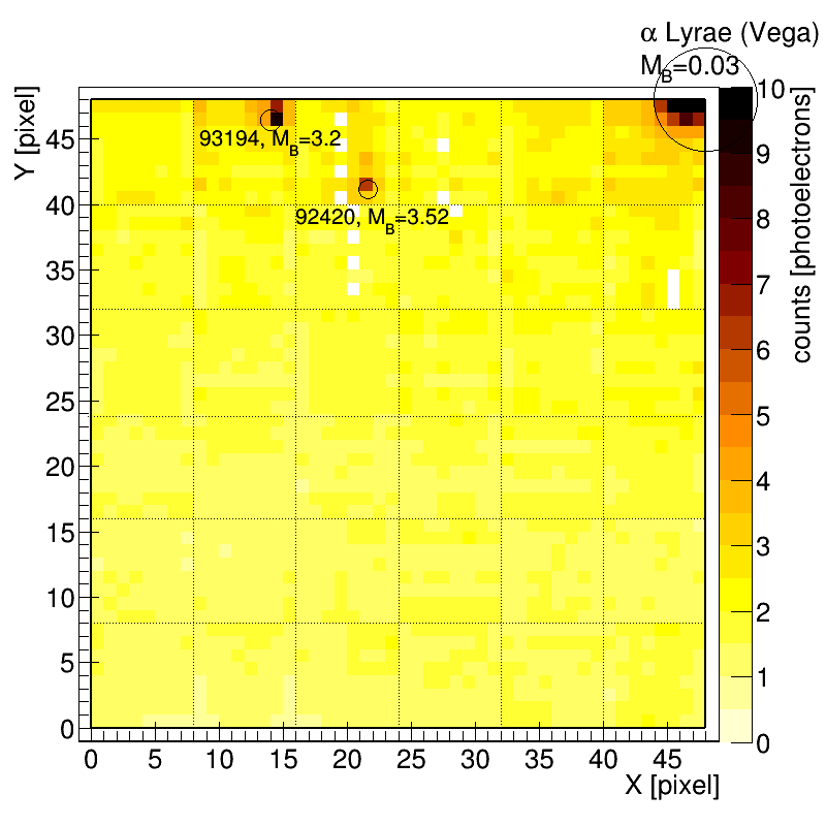}
    \caption{}
    \label{fig:410_starFoV}
  \end{subfigure}
  \hfill    
  \begin{subfigure}[b]{.48\textwidth}
     \includegraphics[width=\linewidth]{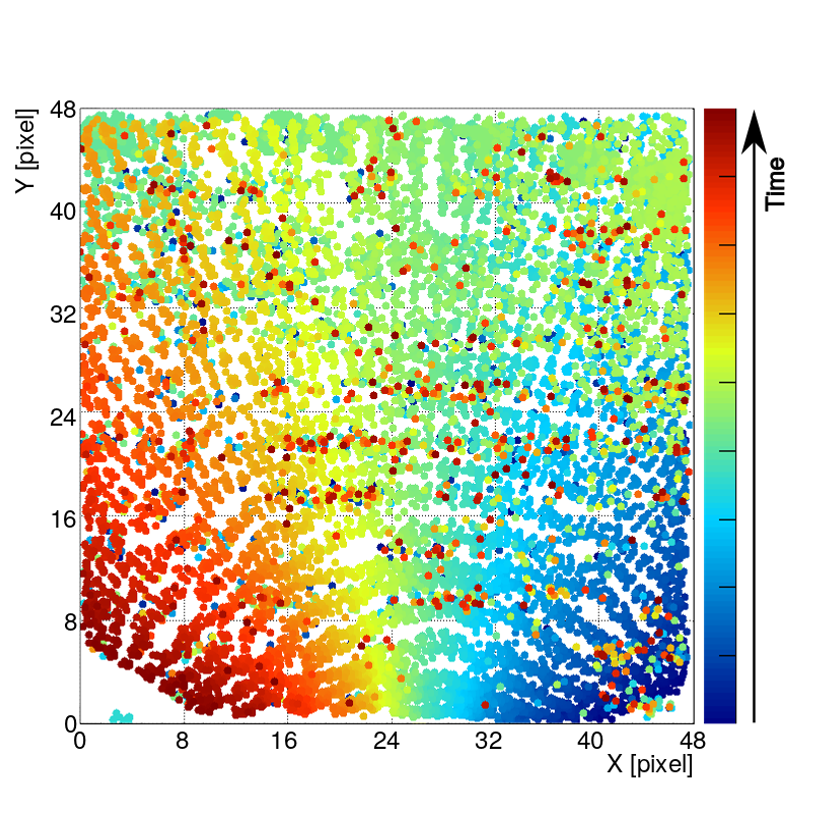}
      \caption{}
      \label{fig:410_laserFoV}
    \end{subfigure}
  \caption{Sample events used to determine the FoV of EUSO-SPB1. (a) Stars within the FoV. Stars from Hipparcos catalog (circles) with a magnitude above 4 are overlaid. (b) Laser sweep in the perpendicular plane to the optical axis. Each dot represents the center of a signal cluster. The red-orange dots are noise signal identified by the trigger algorithm and are not associated with the laser shots. The same is true for the cluster of green dots that appears out-of-sync in the top right corner.}
\end{figure}

An accurate knowledge of the instrument FoV is crucial to estimate the exposure during the flight, and hence to obtain an accurate expected event rate. This made the FoV estimation one of the main goals of our field tests. We measured the FoV with two independent methods. In astronomy, it is standard procedure to measure a telescope FoV based on the transit time of identified bright stars across the FoV. We show an example of the signals of stars in our detector in Fig. \ref{fig:410_starFoV}. Average result based on this method applied to 5 different stars gave us a FoV of \SI[separate-uncertainty = true,multi-part-units=repeat]{11.1(2)}{\degree}. The largest uncertainty for this method is the uncertainty in determining the star position within our camera.

In the second method we used laser tracks swept in \SI{2}{\degree} steps on the
perpendicular plane to the line formed by the EUSO-SPB1 and laser position (Fig. \ref{fig:410_laserFoV}). The source of uncertainties in this method is the laser pointing direction, which we know better than \SI{0.2}{\degree} and the relative position of the GLS prototype and the instrument both of which have only GPS uncertainties (around 5 meters). With this approach we estimated the FoV to \SI[separate-uncertainty = true,multi-part-units=repeat]{11.2(1)}{\degree}. The two methods are consistent within errors but slightly lower than the predicted FoV of \SI{12}{\degree} given by raytrace simulations. Another way to express this FoV is the plate scale which is in our case $\sim$ \SI{0.2}{\degree/pixel}. This already takes into account the dead space between PMTs and ECs.

\subsection{Absolute Calibration with LED}
\label{subsec:AbsCal}
\begin{figure}[h]
  \centering
  \begin{subfigure}{.48\textwidth}
    \includegraphics[width=\linewidth]{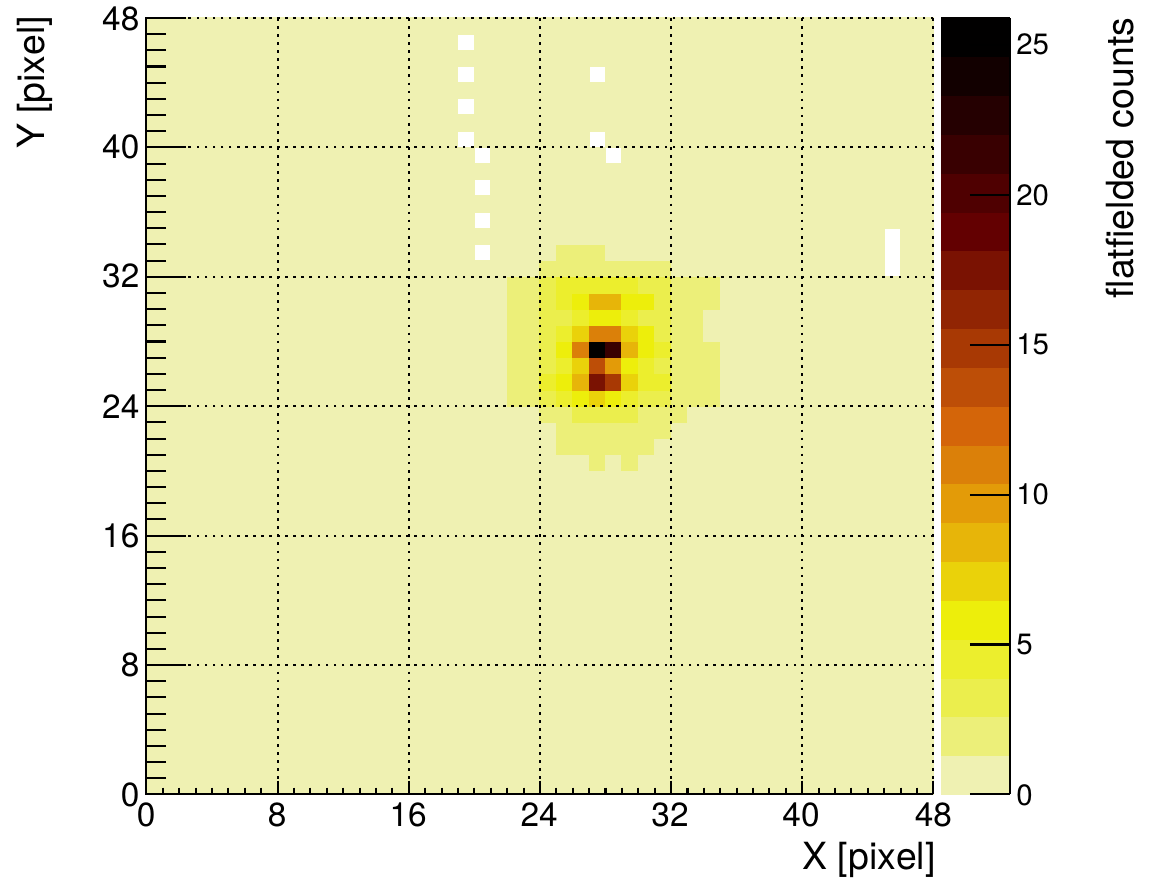}
    \caption{}
    \label{fig:420_LEDExample}
  \end{subfigure}
  \hfill    
  \begin{subfigure}{.46\textwidth}
     \includegraphics[width=\linewidth]{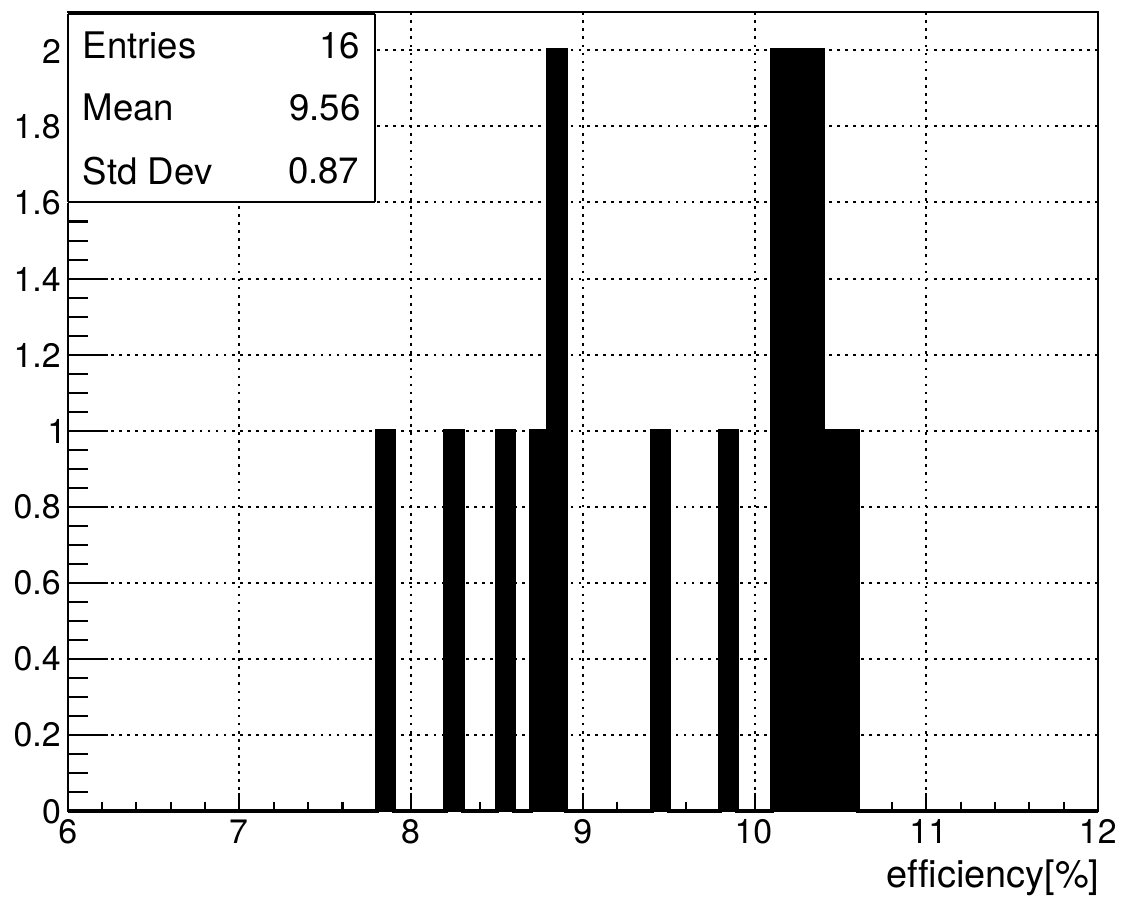}
      \caption{}
      \label{fig:420_AbsEff}
    \end{subfigure}
  \caption{Photometric absolute calibration of EUSO-SPB1 as measured during the Utah field campaign in September 2016 for the two-lens configuration of the instrument. (a) Average over 206 LED pulses after background subtraction and applied flat fielding. (b) Distribution of the absolute efficiencies for all 16 combinations of LED positions and voltages used.}
\end{figure}

An important characteristic of the instrument is the relationship between the incident photon flux at the aperture and the signal (photoelectrons) in the camera. To obtain this relationship we performed an end-to-end absolute photometric calibration in the field. The light source was a \SI{365}{\nano\meter}-LED mounted on a portable radio mast \SI{45}{\meter} away from the detector aperture. A temperature stabilizing circuit inside the LED decreased the temperature effect on the LED  output intensity to less than 4\%. The movable mast allowed us to probe different parts of the PDM by varying the height as well as the lateral position of the mast. The LED was operated in pulse mode with a pulse length of \SI{50}{\micro\second} (equal to 20 GTUs). We performed the measurement for both lens systems (although for the 3-lens system we had to use a different, not temperature stabilized LED). We calculated the calibration constant as the ratio of measured signal after applying flat fielding and subtracting background noise to the number of arriving photons at the aperture. The aperture of the EUSO-SPB1 lens was \SI{0.93}{\square\meter}. We calculated the photon number based on the measured distance from the front lens to the LED and the absolute calibration of the LED's luminosity. The LED voltage used for the analysis was varied between \SIlist{400;1500}{\milli\volt}, providing between 926$\pm$93 and 4156$\pm$419 photons per pulse at the aperture. There are three contributions to the error in the photon number: the dominant one is the uncertainty on the distance between the LED and the EUSO-SPB1 aperture during the field tests, the second and third are the uncertainties on the properties of the LED pulse during the characterization in the lab and on the distance between the LED and the calibrated PMT in the lab. Fig. \ref{fig:420_LEDExample} shows an example LED event after flat fielding and background subtraction. For this example, the number of photons on the aperture was 4156$\pm$419 photons and the sum of the recorded signal was \SI[separate-uncertainty = true,multi-part-units=single]{425(21)}{counts} (within a box of 9 by 9 pixels). The quoted error is the statistical error on the photo-electron count. It has to be stated that this figure does not reflect the actual PSF as our instrument was focused to infinity but the LED was only \SI{45}{\meter} away. In Fig. \ref{fig:420_AbsEff} we show the distribution of the absolute efficiency. Each entry corresponds to a different LED setting and/or location. This way, we estimated the overall system efficiency of EUSO-SPB1 to \SI[separate-uncertainty = true,multi-part-units=single]{0.10(1)}{pe/photon} with two-lens system while the three-lens system's efficiency was found to be lower (\SI[separate-uncertainty = true, multi-part-units=single]{0.04(1)}{pe/photon}).
The overall efficiency of $\sim$10\% was found to be acceptable for EUSO-SPB1 given its relative proximity to the UHECR showers to be detected. For a space-based experiment however, such relatively low efficiency would raise the detection threshold considerably and the overall experiment would not meet its primary scientific objectives.

\subsection{Energy Threshold Estimation}
\label{subsec:EThres}
\begin{figure}[h!] 
	\centering 
	\includegraphics[width =.75\textwidth]{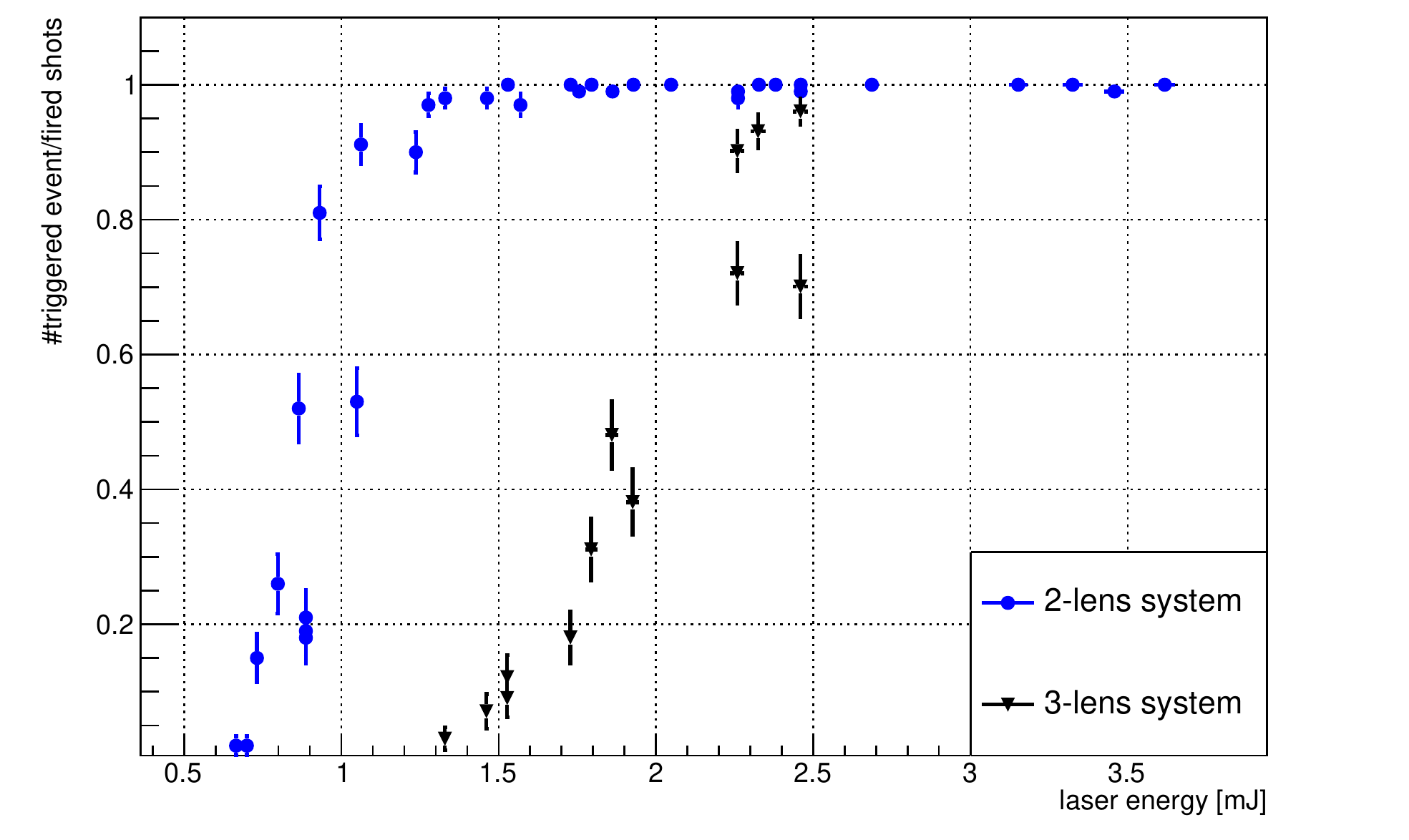}
	\caption{Trigger efficiency of EUSO-SPB1 as a function of laser energy measured during the Utah field campaign in 2016. Blue circles are measurements of the 2-lens system (two nights). Black triangles show the results of the 3-lens configuration (one night).}
	\label{fig:430_TriggEff}
\end{figure}

Another important characteristic of the instrument is the energy threshold of detection of EAS. Once again, we used the portable laser system to perform measurements to estimate this threshold for the onboard trigger. This logic triggers on EAS are based on a signal to noise ratio localized in a cell of 3$\times$3 pixels with the signal having to last for at least 2 GTUs. The threshold is dynamically changed at the MAPMT level. This way, we are able to strongly suppress false positives triggers caused by fluctuations due to nightglow and electronic noise (more details are available in Ref. \cite{SPB1_2018_Battisti_Trigger}). For this study we placed the laser at a distance of \SI{24}{\kilo\meter} in front of the detector. The laser beam was oriented \SI{45}{\degree} away from the detector. We chose these settings as they are comparable to the distance and the most common relative angle at which the EUSO-SPB1 fluorescence telescope could be expected to observe EASs, while looking down on the troposphere during the balloon flight. Using this setup we fired 100 laser shots for each of the 21 laser energy settings (from \SIrange{0.5}{4}{\milli\joule}) and counted the number of triggered events for each setting. The study was performed for both lens configurations to determine which setting configuration had the lower energy threshold hence the higher probability to detect EAS. We show the results in Fig. \ref{fig:430_TriggEff}. We eliminated noise generated triggers and took into account the effect of clouds by analyzing the light profile shapes as well as long exposure pictures of the sky. Clouds can reduce the trigger rate by blocking light. But clouds also can increase the trigger rate: when the laser beam directly hits a cloud more light is scattered increasing the amount of photons scattered in the direction of the detector. We define the energy threshold as the energy at which we have a 50\% trigger efficiency. For the two-lens system that threshold is found to be at \SI[separate-uncertainty = true, multi-part-units=single]{0.94(2)}{\milli\joule} and for the three-lens system at \SI[separate-uncertainty = true, multi-part-units=single]{1.97(3)}{\milli\joule}. This factor of 2 is
consistent with laboratory measurements (see Section \ref{sec:LabTest}) of the relative optical efficiency of the 2-lens and 3-lens configurations. Following these measurements we decided to fly the 2-lens configuration.
Obtaining a 50\% detection threshold equivalency based on scattered light between laser and EAS is desirable for obvious reasons. This can be done with the help of Monte-Carlo simulations although it remains dependent on geometry. Considering a laser pulse observed from an altitude of \SI{33}{\kilo\meter} with a \SI{45}{\degree} zenith angle, we find that a \SI{0.9}{\milli\joule} \SI{355}{\nano\meter} laser pulse produces the same amount of light as a \SI{3.5}{EeV} EAS at maximum under the same conditions.

\subsection{Reconstructed Angular Resolution}
\label{subsec:AngRes}
The arrival direction of an extensive air shower is the first step to estimate the parameters of the primary cosmic-ray that induced it. Therefore, a high angular resolution with a small reconstruction error is essential for any meaningful reconstruction.\\
The analysis has two major steps. First, the pointing direction of the selected pixels are used to find the Shower Detector Plane (SDP). Second, a time fit is performed with the following function:
	\begin{equation}
	\label{eq:3parameter}
	t_{i, expected} = T_0 + \dfrac{R_P}{c} \tan\left(\dfrac{\pi}{4}+\dfrac{\psi_0 - \psi_i}{2}\right),
	\end{equation}
where $\psi_i$ is the pointing direction of each participating pixel projected into the SDP. $T_0$ is the time when the shower front reaches $R_P$, the distance of closest approach. $\psi_0$ is the angle from horizontal to $R_p$. A $\chi^2$ minimization is used to find the best time fit to the data. In the field test, it is possible to reduce the fit function to 2 parameters using the known position of the laser.\\
To create the data set for this analysis we used laser shots varying the angle, called $\alpha$, between the laser beam and the optical axis (tilted upwards by \SI{7.8}{\degree} from horizontal) of the instrument in \SI{5}{\degree} increments between \SI{13}{\degree} and \SI{88}{\degree}(\SI{0}{\degree} means the laser beam is perpendicular to the optical axis and positive values indicate a laser pointing direction away from the instrument). The laser energy was \SI{2}{\milli\joule}. Note that for an expected angle above $\sim$\SI{70}{\degree}, we are not able to identify the tracks as the signal becomes too dim.
We define the reconstruction angular resolution ($\alpha_{\text{rec}}^{\text{err}}$) as the error on the reconstructed angle ($\alpha_{\text{rec}}$).
\begin{figure}[h]
  \centering
  \includegraphics[width=.75\linewidth]{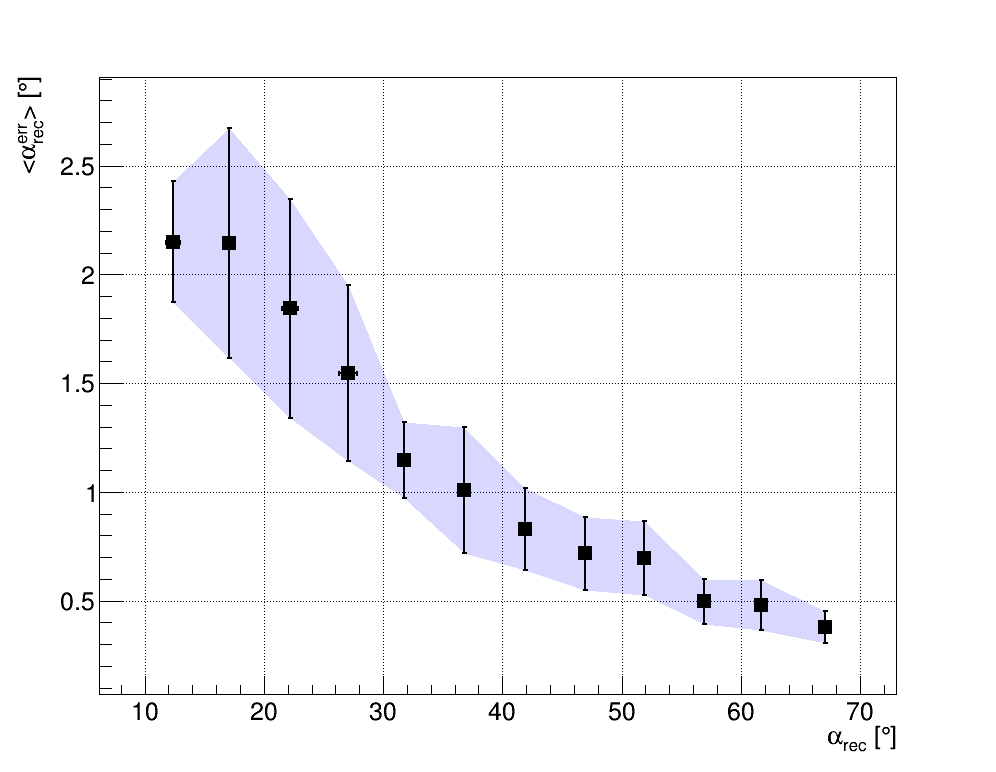}   
  \caption{Mean Angular resolution ($<\alpha_{\text{rec}}^{\text{err}}>$) for EUSO-SPB1 as a function of reconstructed angle $\alpha_{\text{rec}}$. The error shown represents the std. deviation on the error of the reconstructed angle $\alpha_{rec}^{err}$.}
  \label{fig:440_angularRes}
\end{figure}
The angular resolution is, in the worst case ($\alpha$=\SI{13}{\degree}, close to vertical laser pointing direction), \SI{2.2}{\degree} and decreases to \SI{0.4}{\degree} at a reconstructed angle of \SI{68}{\degree}. This is expected because for more tilted trajectory the signal persists for longer time within the FoV providing a better fit.\\
We also compared the reconstructed angle ($\alpha_{\text{rec}}$) with the expected angle ($\alpha_{\text{exp}}$) which again is the angle between the laser pointing direction and the instrument optical axis tilted by \SI{7.8}{\degree} from horizontal and found that we have a systematic shift to lower angles by around \SI{0.9}{\degree} which could be caused by misalignment of the laser tilt or the telescope.

\section{Summary}
Before the launch of EUSO-SPB1 in April 2017, we carried out an extensive and detailed characterization of the detector, starting out with measuring the single components in the laboratory. These test showed an efficiency of the PDM of  \SI[separate-uncertainty = true,multi-part-units=single,allow-number-unit-breaks]{0.30(3)}{pe/photon} for a wavelength of \SI{378}{\nano\meter}. For the two-lens flight configuration, we found that 31 $\pm$ 2\% of photons arriving at the aperture are contained in a PSF of \SI{9}{\mm} diameter at the PDM. The flat field measurements made in Wanaka (NZ) before launch allowed us to equalize the pixel sensitivity within $\sim$ 6\%.\\
In a second step, we performed an end-to-end test of the instrument setup at the TA site in the Utah desert. 
We conducted a photometric calibration of the assembled instrument resulting in \SI[separate-uncertainty = true,multi-part-units=single]{0.10(1)}{pe/photon}. This is comparable to the combined laboratory results. We estimated the instrument's FoV of \SI[separate-uncertainty = true,multi-part-units=repeat]{11.1(2)}{\degree}. We found that a 50\% trigger efficiency is reached by the scattered light flux of a \SI{0.9}{\milli\joule} laser beam. Finally, we estimated the angular resolution ($\alpha_{\text{rec}}^{\text{err}}$) to be better than \SI{2.2}{\degree} for the conditions during the field tests while using the laser position as a constraint.\\
The collaboration is preparing a follow-up mission, EUSO-SPB2 \cite{EUSO-SPB2}
with an anticipated launch date of March 2022. This new instrument will be equipped with two telescopes using Schmidt optics, expecting a higher light collection efficiency. One is optimized for the detection of EAS from UHECR. The second one is optimized for
the Cherenkov light detection of tau-neutrino induced showers by using Silicon-Photomultipliers with a very fast readout electronics. Even though these telescopes are using different optics and one at least different type of electronics, we will perform similar test based on th experience gained in EUSO-SPB1.
\section*{Acknowledgment}
This work was partially supported  
by NASA grants NNX13AH54G, NNX13AH55G, NNX13AH53G, NNX13AH52G, NNX16AG27G, 80NSSC18K0246, 80NSSC18K0477, 80NSSC18K0473, 80NSSC18K0464, 
the French Space Agency (CNES),
the Italian Space Agency through the ASI INFN agreement n. 2017-8-H.0,
the Italian Ministry of Foreign Affairs and International Cooperation,
the Basic Science Interdisciplinary Research Projects of RIKEN and JSPS KAKENHI Grant (22340063, 23340081, and 24244042), 
by the Deutsches Zentrum f\"ur Luft- und Raumfahrt,
the Helmholtz Alliance for Astroparticle Physics funded by the Initiative and Networking Fund 
of the Helmholtz Association (Germany),
by the Mexican funding agencies PAPIIT-UNAM, CONACyT and the Mexican Space Agency (AEM) 
and by the National Science Centre in Poland grant (2017/27/B/ST9/02162).
We acknowledge the Telescope Array Collaboration for the use of their facilities in Utah as well as the Goddard Space Flight Center for the loan of the mirror used for the lens test. We also acknowledge the invaluable contributions of the administrative and technical staffs at our home institutions.

\bibliographystyle{spphys}       
\bibliography{Reference.bib}   

\end{document}